\documentclass[a4paper,11pt]{article}
\usepackage{pos}
\usepackage{amsfonts,amsmath,hyperref,url, color}
\usepackage{bm, bbm}
\usepackage{graphicx}
\usepackage{mathtools}

\usepackage[utf8]{inputenc}
\usepackage[english]{babel}

\usepackage{hyperref}
\usepackage{caption,adjustbox}
\usepackage{multirow}
\usepackage{cleveref}
\usepackage{fontawesome}
\usepackage{comment}
\usepackage{graphicx}
\usepackage{float}
\usepackage{bbding}
\usepackage{xcolor}
\usepackage[arrowdel]{physics}
\usepackage{tensor}
\usepackage{cancel}
\usepackage{tikz}
\usepackage{cleveref}

\usepackage{listings} 

\usepackage[title,toc,titletoc]{appendix}
\usepackage{titlesec}
\usepackage{tocloft}

\newcommand{\be}{\begin{equation}}
\newcommand{\ben}{\begin{equation*}}
\newcommand{\ee}{\end{equation}}
\newcommand{\een}{\end{equation*}}
\newcommand{\ba}{\begin{eqnarray}}
\newcommand{\ea}{\end{eqnarray}}

\title{Entanglement entropy and horizon temperature in
conformal quantum mechanics}

\author*[a,b]{Michele Arzano}
\author[a,b]{Alessandra D'Alise}
\author[a,b]{Domenico Frattulillo}

\emailAdd{michele.arzano@na.infn.it}

\affiliation[a]{Dipartimento di Fisica ``E. Pancini", Universit\`a di Napoli Federico II, I-80125 Napoli, Italy\\}
\affiliation[b]{INFN, Sezione di Napoli,\\ Complesso Universitario di Monte S. Angelo,\\
Via Cintia Edificio 6, 80126 Napoli, Italy}

\emailAdd{alessandra.dalise@unina.it}

\emailAdd{domenico.frattulillo@unina.it}

\abstract{The generators of radial conformal symmetries in Minkowski space-time can be put in correspondence with generators of time evolution in conformal quantum mechanics. Within this correspondence we show that in conformal quantum mechanics the state corresponding to the inertial vacuum for a conformally invariant field in Minkowski space-time has the structure of a thermofield double. The latter is built from a bipartite ”vacuum state” corresponding to the ground state of the generators of hyperbolic time evolution. These can evolve states only within a portion of the time domain. When such generators correspond to conformal Killing vectors mapping a causal diamond in itself and generators of dilations, the temperature of the thermofield double reproduces, respectively, the diamond temperature and the Milne temperature found for massless fields in Minkowski space-time. Moreover, we compute the entanglement entropy associated to the thermofield double states obtaining a UV divergent logarithmic behaviour akin to known results in two-dimensional conformal field theory where the entangling boundary is point-like.}

\FullConference{%
  Workshop on Noncommutative and Generalized Geometry in String Theory, Gauge Theory and Related Physical Models\\
 September 22, 2023 \\
Corfu Summer Institute, Corfu, Greece\\}

\begin{document}
\maketitle

\section{Introduction}
In quantum field theory the definition of vacuum state relies on the choice of a time-like Killing vector to which one associates a notion of positive frequency for the modes of the field. In this sense, whether such state is devoid of particles depends on the way one observer measures time. This seemingly straightforward yet profoundly significant insight was first being appreciated in the 1970s (see e.g. the nice essay by P.C.W. Davies \cite{Davies:1984rk}) and it is at the basis of the existence of the temperature which certain classes of observeres associate to a Killing horizon. In ordinary Minkowski space-time, the well-known example is the Unruh temperature: uniformly accelerating observers perceive what is the vacuum state for inertial observers as a thermal state at a temperature proportional to the modulus of their four-acceleration \cite{Fulling:1972md,Unruh:1976db}. In this example, the vacuum state for inertial observers, which define positive frequency modes with respect to globally time-like Killing vector (the generator of translations in inertial time), appears as a thermal state to accelerated observers whose notion of positive frequency is determined by the generator of boosts which is time-like only within the so-called Rindler wedges, the regions of Minkowski space-time determined by the causal complement of a light-cone.
The same ambiguity in the definition of a quantum field's vacuum state is ultimately responsible for celebrated effects in semiclassical gravity as cosmological particle creation \cite{Parker:1968mv} and particle creation by black holes \cite{Hawking:1975vcx}.

Much less appreciated is the fact that for massless fields, for which the invariance group of the field equation is enlarged from the Poincaré to the conformal group, time-like {\it conformal Killing vectors}, in addition to usual time-like Killing vectors, can be used to characterize positive frequency modes of the field and thus to define associated vacuum states. One could for example use the generator of dilations in Minkowski space-time to define a decomposition of the field in modes which have positive and negative frequencies with respect to conformal time in a Milne universe (a flat expanding (contracting) FRLW space-time which can be identified with the future (past) cone in Minkowski space-time). The usual inertial vacuum state of the field appears as a thermal state populated of such Milne modes \cite{Higuchi:2017gcd,Wald:2019ygd}. Similarly, as discussed in detail below, a combination of special conformal tansformations and time translation generates time evolution which preservers a {\it causal diamond}, the intersection of a past and future cone in Minkowski space-time. Observers whose worldlines are orbits of such conformal Killing vector, like Milne observers, see the inertial vacuum as a state populated by positive frequency excitations with respect to the {\it diamond time} determined by the conformal Killing vector \cite{Martinetti:2002sz,Su:2015oys,Perez:2023pcl}. 

In \cite{Arzano:2020thh}, it was observed that radial conformal Killing vectors in Minkowski space-time, which include the vectors defining Milne and diamond time, once restricted to light cones, are formally identical to the generators of time evolution in conformal quantum mechanics, a quantum mechanical model with inverse square potential first studied in detail in \cite{deAlfaro:1976vlx}. Conformal quantum mechanics can be seen as a conformally invariant $0+1$-dimensional field theory. Any generator of the group of conformal transformations of the real line can be used as a generator of time evolution of the theory. In the same way for a quantum field in Minkowski space-time the vacuum state defined by a (conformal) Killing vector that is globally time-like appears thermal to observers whose time evolution is determined by a (conformal) Killing vector which is not globally time-like, one might expect a similar effect to be present in conformal quantum mechanics. As shown in \cite{Arzano:2020thh,Arzano:2021cjm}, and as we will review in this contribution, this is indeed the case.\\

The significance of this observation is twofold. On one side it shows how in an extremely simplified one-dimensional model one can reproduce effects analogous to their semiclassical gravity counterparts and thus provide further insights on their fundamental aspects. In particular, it evidences how such effects are ultimately related to the possibility of defining alternative notions of time evolution which cover only part of the geometric domain of the theory. On the other hand, as observed in \cite{Arzano:2020thh}, since the two-point function of conformal quantum mechanics seen as a CFT$_1$ \cite{Chamon:2011xk,Jackiw:2012ur} can reproduce, for specific choices of the basis states of the theory, the two-point function of a massless scalar field in Minkowski space-time evaluated along the worldline of Milne and diamond observers, its thermal nature can be used as an alternative, group-theoretical derivation of the temperatures that these observers associate to the inertial vacuum state.

The other crucial characteristic of the inertial vacuum state, when seen from the perspective of excitations with frequency defined by non-globally time-like (conformal) Killing vectors, is its entangled nature. In fact, the thermal character of this state seen from observers whose time evolution is restricted to a portion of the whole space-time (e.g. in the case of Minkoski space-time the Rindler wedge, the Milne universe or a causal diamond) is due to the restriction of this highly entangled state to a portion of space-time which requires tracing over the modes not having support on that region \cite{Higuchi:2017gcd,Wald:2019ygd,Su:2015oys,Olson:2010jy}. As we review in this contribution, the same pattern occurs in the case of conformal quantum mechanics where the state which plays the analogue role of the inertial vacuum exhibits the structure of a {\it thermofield double state}. We will calculate the entanglement entropy associated to such state and show how it  exhibits the same logarithmic divergence characterizing field theories in which the boundary of the entangling region is point-like. Seen as a quantum field theory, conformal quantum mechanics thus likely provides the simplest field-theoretic model where an analytic calculation of entanglement entropy
associated to a geometric partition is possible.

\section{Radial conformal Killing vectors in Minkowski space-time}\label{rkv}

We start from the metric of Minkowski space-time written in spherical coordinates 
\be
d s^2 = -d t^2 + d r^2 + r^2 (d \theta^2 + \sin^2 \theta\,  d\phi^2)\,.
\ee
Given a radial vector field 
\begin{equation}\label{rvf}
    \xi= \xi_t(t,r,\theta,\phi) \pdv{}{t}+ \xi_r(t,r,\theta,\phi) \pdv{}{r}
\end{equation}
we say that it is a  {\it conformal Killing vector} if
\be
\mathcal{L}_{\xi} g_{\mu\nu} \propto g_{\mu\nu}\,.
\ee
From this condition one can obtain \cite{RCM} the general form of a radial conformal Killing vector given by
\be\label{xi11}
\xi = \left(a(t^2+r^2)+b t +c\right)\, \partial_t + r (2 a t + b)\, \partial_r
\ee
with $a,b,c$ real constants. The conformal Killing vector above can be written as 
\be
\xi = a K_0 + b D_0 + c P_0\,,
\ee
where $P_0$, $D_0$ and $K_0$ generate, respectively, time translations, dilations and special conformal transformations
\be
P_0  =  \partial_t\,,\qquad D_0  =  r\, \partial_r + t\, \partial_t\,,\qquad K_0  =  2 t r\, \partial_r + (t^2+r^2)\, \partial_t \, .
\ee
These generators close the $\mathfrak{sl}(2,\mathbb{R})$ Lie algebra 
\be
[P_0,D_0]= P_0\,,\qquad [K_0,D_0]= - K_0\,,\qquad [P_0,K_0]= 2 D_0\,.
\ee
Introducing null coordinates $u=t-r$ and $v=t+r$, the radial conformal Killing vector \eqref{xi11} divides into two identical $u$- and $v$-dependent terms 
\begin{equation}
    \xi=  \left(a u^2+b u +c\right)\, \partial_u+\left(a v^2+b v +c\right)\, \partial_v\ .
\end{equation}
The causal structure of the conformal Killing vectors changes according to the values of the real constants $a,b$ and $c$ \cite{RCM}. When $a=0$ and $b = 0$ the conformal Killing vectors are globally time-like. When $a=0$ and $b\neq 0$ they are null on the light-cone originating from the point $(t= -c/b,r=0)$, time-like inside it and space-like outside. When $a\neq 0$ the domains of causality of the Killing vector are determined by the sign of the determinant $\Delta =b^2-4 ac$:
\begin{itemize}
    \item $\Delta<0$: $\xi$ is time-like everywhere.
    \item $\Delta=0$: $\xi$ is time-like everywhere but on the light-cones emanating from $(t=-b/(2a),r=0)$ where it is null.
    \item $\Delta>0$: $\xi$ is null on the light-cones originating from the points $(t_\pm,r=0)$ (where $t_\pm=(-b\pm \sqrt{\Delta})/2 a$), it is time-like inside and outside of both light-cones and space-like everywhere else. 
\end{itemize} 

When the conformal Killing vector is time-like, we can define an observer having four-velocity $U^\mu=\exp{-\varphi}\xi^\mu$ where $\xi^\mu\xi_\mu=\exp{2\varphi}$, we call this a \emph{conformal Killing observer} \cite{dyer1979conformal}\footnote{The scalar function $\varphi$ is defined as $\mathcal{L}_\xi\ g_{\mu\nu}=2\varphi\ g_{\mu\nu}$.}. In the regions of space-time where the conformal Killing vector is not time-like such an observer cannot exist and one defines the concept of a conformal stationary limit surface as the boundary where a conformal Killing vector becomes light-like. Such boundary is a null surface and defines a \emph{conformal Killing horizon} when it is also a geodesic hypersurface.

Key to our discussion is the observation that radial conformal Killing vectors (Eq. \eqref{xi11}) restricted to $r=0$ and $u=const$ or $v=const$  assume the form of the generator of a conformal transformation of the real line
\be\label{xikv}
\xi =\left( a\, t^2 + b\, t + c\, \right) \partial_t\,.
\ee
This provides the bridge for the correspondence between orbits of conformal Killing vectors in Mikowski space-time and time evolution in conformal quantum mechanics which we discuss in detail in \cref{Rstates}. In preparation for that, below we discuss how the worldlines of observers in a Milne universe and in a causal diamond in Minkowski space-time can be described by orbits of radial conformal Killing vectors.

\section{Observers worldlines and orbits of (conformal) Killing vectors}
Radial conformal Killing vectors divide Minkowski space-time into causal domains separated by light-like surfaces that are conformal
Killing horizons, the latter having the property that the conformal Killing vector field is orthogonal to them \cite{Nielsen:2017hxt}.  We now show how orbits of the radial conformal Killing vectors we discussed in the previous section describe worldlines of observers restricted to the time-like regions of these causal domains.
For simplicity, we restrict to the $1+1$-dimensional case. The line element of Minkowski space-time in cartesian coordinates is
\begin{equation}
    ds^2=-dt^2+dx^2\,.
\end{equation}
We recall that a set of coordinates is {\it adapted} to the worldline of a class of observers determined by integral lines of a time-like vector if such a vector can be expressed as a derivative with respect to the time coordinate and the space coordinates stay constant along the worldline. Cartesian coordinates are {\it adapted} to static inertial observers. Such observers are those whose wordlines are integral lines of the (globally) time-like Killing vector $P_0=\partial_t=(1,0)$ and whose four-velocity is parallel to it
\begin{equation}
    u^{\mu}_s||P_0\,,
\end{equation}
(generally inertial observers can have four-velocities which are not parallel to $P_0$, see below).
 In fact, when using cartesian coordinates we have indeed that $P_0=\partial_t$ and, for static observers, such coordinate coincides with the {\it proper time} along the worldline i.e. the length
\begin{equation}
    \tau = \int dt \sqrt{-g_{\mu\nu} \frac{dx^{\mu}}{dt}\,\frac{dx^{\nu}}{dt}}\,.
\end{equation}
Moreover the spatial coordinate $x$ stays constant along worldlines of static inertial observers.\\
\subsection{Rindler observer}
Let us now consider the case of {\it Rindler coordinates}. These are coordinates adapted to an  observer moving along a trajectory whose acceleration four-vector has a constant modulus. The worldline of such an observer is the integral line of the boost Killing vector $N_0 = x\partial_t+t\partial_x=(x,t)$. By definition, integral lines of a Killing vector have the Killing vector itself as tangent vectors and thus the four-velocity of observers coincides with the Killing vector as well. From the normalization of the four-velocity we generally have
\begin{equation}\label{ueq1}
    -(u^0)^2+(u^1)^2 = -1\ .
\end{equation}

A way to find integral curves is simply to find a curve $\Gamma_{N_0}(\lambda)=(t(\lambda),x(\lambda))$ such that its tangent vector coincides with the vector $N_0$ i.e.
\begin{align}
 u^0= \frac{dt(\lambda)}{d \lambda} & = x\\
 u^1= \frac{dx(\lambda)}{d \lambda} & = t
\end{align}
whose normalized solution is
\begin{align}
    u^0 & = \cosh(\lambda) \\
    u^1 & = \sinh(\lambda) \,,
\end{align}
which can be rewritten in terms of an integration constant $\alpha$ determining the intersection of the worldline with the $x$-axis and a dimensionful time coordinate $\sigma=\alpha \lambda$
\begin{align}\label{Ri1}
    t & = \alpha \sinh\left(\frac{\sigma}{\alpha}\right) \\ \label{Ri11}
    x & = \alpha \cosh\left(\frac{\sigma}{\alpha}\right) \,.
\end{align}
Rindler observers move along hyperbolae and thus, to define adapted coordinates to these integral lines, one needs to choose another spatial coordinate which stays constant along the trajectory instead of $x$. Since in the limit $\alpha \rightarrow 0$ the trajectory tends to the light-cone $x=|t|$ the worldlines are restricted to the right Rindler wedge $x>|t|$. All other worldlines will intersect the $x$-axis everywhere from $x=0$ to $x=\infty$ and they can be written in terms of the action of a dilation on the worldline \eqref{Ri1} and \eqref{Ri11}
\begin{align}\label{Ri2}
    t   & = \alpha\, e^{\frac{\chi}{\alpha}} \sinh\left(\frac{\sigma}{\alpha}\right) \\ \label{Ri22}
    x   & = \alpha\, e^{\frac{\chi}{\alpha}} \cosh\left(\frac{\sigma}{\alpha}\right) \,,
\end{align}
where the dilation parameter $\chi\in(-\infty,\infty)$ is the Rindler spatial coordinate which stays constant along a given Rindler trajectory. The Minkowski metric written in Rindler coordinates reads
\begin{equation}\label{rindlerm2}
    \dd s^2= e^{2 \frac{\chi}{\alpha}} \left(-\dd \sigma^2+\dd \chi^2\right)
\end{equation}
from which one can read the proper time to be
\begin{equation}
    \tau=e^{ \frac{\chi}{\alpha}} \sigma\ .
\end{equation}
Notice that  there is only one Rindler worldline whose proper time coincides with the Rindler coordinate time: the one at the origin of the spatial axis of the Rindler coordinates chosen, $\chi=0$.

\subsection{Milne observer}
We now define {\it Milne observers} as the ones whose worldlines are integral curves of the dilation {\it conformal Killing vector} $D_0 = t\partial_t + x \partial_x = (t,x)$. 
As before, one can find integral curves by searching for a curve $\Gamma_{D_0}(\lambda)=(t(\lambda),x(\lambda))$ such that its tangent vector coincides with the vector $D_0$ i.e.
\begin{align}
  \frac{dt(\lambda)}{d \lambda} & = t\\
  \frac{dx(\lambda)}{d \lambda} & = x\,.
\end{align}
A trivial solution is $(t=0, x=0)$. A general solution is given by
\begin{align}\label{Mi2}
    t & = t_0\, e^{\lambda} \\
    x & = x_0\, e^{\lambda} \,,
\end{align}
with initial conditions $(t(\lambda=0) = t_0, x(\lambda=0) = x_0)$. We see that integral lines of $D_0$ are straight lines through the origin
\begin{equation}\label{milnew}
    x = \frac{x_0}{t_0}\ t=\frac{t}{\omega}
\end{equation}
and are time-like for $\abs{\omega} > 1$, space-like for $\abs{\omega} < 1$ and null when $\abs{\omega}=1$. We can now introduce a length scale $\alpha$ and write, for time-like, future-oriented integral lines
\begin{align}\label{Mi3}
    t & = \alpha \cosh\zeta\, e^{\lambda} \\
    x & = \alpha \sinh\zeta\, e^{\lambda} \,,
\end{align}
with the boost-like parameter $\zeta\in(-\infty,\infty)$. 
We introduce a time coordinate $\sigma = \alpha \lambda$ and a space coordinate $\chi = \alpha \zeta$ \footnote{Notice that if we had introduced different length scales $\beta_{\sigma}, \beta_{\chi} \neq\alpha$ to convert the dimensionless parameters $\lambda$ and $\zeta$ into a time and space coordinates we would have unpleasant factors $(\frac{\alpha}{\beta_{\sigma}})^2$ and $(\frac{\alpha}{\beta_{\chi}})^2$ appearing on the right-hand side of \eqref{milneme2}.} and define Milne coordinates in the future cone of Minkowski space-time as the set $(\sigma,\chi)$ 
\begin{align}\label{Mi31}
    t' & = \alpha\, e^{\frac{\sigma}{\alpha}} \cosh\left(\frac{\chi}{\alpha}\right)  \\ \label{Mi32}
    x' & = \alpha\, e^{\frac{\sigma}{\alpha}} \sinh\left(\frac{\chi}{\alpha}\right)  \,.
\end{align}

Such coordinates are adapted to the integral lines of the dilation conformal Killing vector $D_0$ and indeed one can verify that $D_0 = \alpha \partial_\sigma$ and the coordinate $\chi$ stays constant along such integral lines. We can also verify that, in these coordinates, the boost generator $N_0=x\partial_t+t\partial_x$ reads $N_0=\alpha \partial_\chi$ connecting therefore worldlines of different Milne observers. 

 We can write the Minkowski metric in Milne coordinates as 
\begin{equation}\label{milneme2}
    \dd s^2= e^{2 \frac{\sigma}{\alpha}} \left(-\dd \sigma^2+\dd \chi^2\right)\, ,
\end{equation} 
from which we see that along such integral lines the Milne time $\sigma$ is related to proper time $\tau$ by
\begin{equation}
    \tau = \alpha e^{\frac{\sigma}{\alpha}}\ ,
\end{equation}
where $\tau>0$. Notice that the proper time $\tau$ and the adapted time $\sigma$ {\it can never coincide} along Milne observers worldlines.\footnote{Observe that in literature (see for instance \cite{Mukhanov:2005sc}) the metric of Milne space-time can be found written in terms of a time coordinate $\Bar{t}$ as
\begin{equation}\label{milneme3}
    \dd s^2= -\dd \Bar{t}^2+ \Bar{t}^2\, \dd \chi^2\, ,
\end{equation}
which has the form of a FLRW metric with linear expansion factor $a(\bar{t})=\bar{t}$. These are {\it comoving coordinates} for the Milne universe with $\bar{t}$ being the cosmological time that coincides with the proper time $\tau$ when $\chi$ is constant.}

\subsection{Diamond observer}\label{diam}
Observers whose worldlines are integral curves of the conformal Killing vector $S_0$
\begin{equation}
    S_0=\frac{1}{2\alpha}\left((\alpha^2-t^2-x^2)\partial_t -2tx \partial_x\right) = \left(\frac{1}{2\alpha}(\alpha^2-t^2-x^2),\frac{1}{\alpha}(-tx)\right)\ 
\end{equation}
are called {\it diamond observers}. These observers are confined in the portion of Minkowski space-time defined by the inequality $|t|+|x|<\alpha$. This region is called causal diamond and it is defined by the intersection of past and future light cones of two events. It corresponds to the causal domain of an observer with a finite lifetime $2\alpha$.  To find the integral curves, let us  recall that on the real line the conformal transformation
 \begin{equation}\label{confmsd}
     t' =\frac{\alpha(t+\alpha)}{\alpha-t}\,,
 \end{equation}
 maps the operator $S_0 = \partial_{t}$ into  $D_0 = \partial_{ t'}$ (see \cite{Arzano:2021cjm} for details). 
 
In light-cone coordinates $u=t-x,v=t+x$
we have 
\begin{align}\label{duv}
    D_0&= u\partial_{u}+v\partial_{v}\\\label{suv}
    S_0 &=\left(\frac{\alpha}{2}-\frac{u^{ 2}}{2\alpha}\right)\partial_{u}+\left(\frac{\alpha}{2}-\frac{v^{2}}{2\alpha}\right)\partial_{v}\ .
\end{align}
As it can be easily checked, \eqref{duv} admits an expression similar to \eqref{suv} when applying the map \eqref{confmsd} separately to $u$ and $v$
\begin{align}\label{cmpuv1}
    u' &=\frac{\alpha(u +\alpha)}{\alpha-u }\\\label{cmpuv2}
    v' &=\frac{\alpha(v +\alpha)}{\alpha-v }
\end{align}
namely:
\begin{equation}
    D_0 =u'\partial_{u'} +  v' \partial_{v'}= \left(\frac{\alpha}{2}-\frac{u^2 }{2\alpha}\right)\partial_{u }+\left(\frac{\alpha}{2}-\frac{v^2 }{2\alpha}\right)\partial_{v }\ .
\end{equation}
From these relations, we can also obtain the maps connecting Milne and diamond observers in cartesian coordinates which read:
\begin{align}\label{md1}
t'&=\frac{\alpha(-t^2 +x^2 +\alpha^2)}{t^2 -x^2 +\alpha^2-2\alpha t }\\ \label{md2}
x'&=\frac{2\alpha^2 x }{t^2 -x^2 +\alpha^2-2\alpha t }
\end{align}
Therefore, in light-cone coordinates, the integral curves of $S_0$ can be obtained by from the integral curves of $D_0$ using the maps \eqref{cmpuv1} and \eqref{cmpuv2}. Specifically, we rewrite the integral curves of $D_0$ in \eqref{milnew} in light-cone coordinates
\begin{equation}
    \omega(v'-u')=u'+v'
\end{equation}
and apply \eqref{cmpuv1} and \eqref{cmpuv2} obtaining
\begin{equation}
   \alpha\omega (v -u )=\alpha^2-u  v 
\end{equation}
which, written in cartesian coordinates, reads as
\begin{equation}\label{icd}
    t^{ 2} -(x -\alpha\omega)^2=\alpha^2(1-\omega^2)\ .
\end{equation}
These integral curves are time-like when $|\omega |> 1$ (inside the causal diamond),  space-like when $|\omega |< 1$ (outside the causal diamond) and null when $|\omega|=1$ (boundary of the causal diamond).\\
Following the same procedure, rewriting \eqref{Mi31} and \eqref{Mi32} in light-cone coordinates and applying \eqref{cmpuv1} and \eqref{cmpuv2} to them, we define the diamond coordinates as the set $(\sigma,\chi)$
\begin{align}\label{diamantecoor1}
    t &=\alpha \frac{\sinh \left(\frac{\sigma }{\alpha}\right)}{\cosh \left(\frac{\chi }{\alpha}\right)+\cosh \left(\frac{\sigma }{\alpha}\right)}\\\label{diamantecoor2}
    x &=\alpha \frac{\sinh \left(\frac{\chi }{\alpha}\right)}{\cosh \left(\frac{\chi }{\alpha}\right)+\cosh \left(\frac{\sigma }{\alpha}\right)}\ .
\end{align}
Notice that the parameter $\omega$ using \eqref{diamantecoor1} and\eqref{diamantecoor2} can be written as
\begin{equation}
    \omega=\frac{1}{\tanh{\left(\frac{\chi}{\alpha}\right)}}
\end{equation}
and it is related to the acceleration of a  diamond observer by
\begin{equation}
    a(\omega)=\frac{1}{\alpha\sqrt{\omega^2-1}}\ .
\end{equation}
We can also apply the map \eqref{confmsd} to 
the boost generator
\begin{equation}
   N_0= x\partial_t+t\partial_x\, ,
\end{equation}
 to obtain the generator $M_0$ that connects worldlines of different diamond observers
\begin{equation}
    M_0=-\frac{xt}{\alpha}\partial_t+\frac{1}{2\alpha}\left(\alpha^2-x^2-t^2\right)\partial_x=\partial_\chi\, .
\end{equation}
Notice that the generator $M_0$ in adapted coordinates acts as a translation in the spatial coordinate $\chi$.
In terms of diamond coordinates, the line-element in Minkowski space-time reads
\begin{equation}\label{diamantemetric}
    \dd s^2= \frac{1}{\left(\cosh \left(\frac{\chi }{\alpha}\right)+\cosh \left(\frac{\sigma }{\alpha}\right)\right)^2}\left(\dd \sigma^2-\dd \chi^2\right)\,,
\end{equation}
showing the conformal relation between the diamond space-time and Minkowski space-time. From \eqref{diamantemetric} we see that along such integral lines the diamond time $\sigma$ is related to proper time $\tau$ by

 \begin{equation}
     \tau=\alpha\  \text{csch} \left(\frac{\chi}{\alpha} \right) \left[\log \left(\cosh \left(\frac{\chi +\sigma }{2 \alpha }\right)\right)-\log \left(\cosh  \left(\frac{\chi -\sigma }{2 \alpha }\right)\right)\right]\, ,
\end{equation}
that, as in Milne space-time, never coincides with the adapted time $\sigma$.\\ Let us observe that \eqref{diamantecoor1} and \eqref{diamantecoor2} can also be obtained by applying the following conformal mapping \cite{hislop1982modular, Martinetti:2008ja}
\begin{align}\label{dr1}
    t'  &= \frac{2 t   \alpha^2}{(x  +\alpha)^2-t^2  } \\ \label{dr2}
    x' &= \frac{\alpha(t^2  -x^2  +\alpha^2)}{(x  +\alpha)^2-t^2  }
\end{align}
on Rindler cartesian coordinates \eqref{Ri2} and \eqref{Ri22}.
Finally, let us also observe that combining the maps \eqref{dr1} and \eqref{dr2} that connect a diamond observer with a Rindler one, with \eqref{md1} and \eqref{md2} which connect  Milne and diamond observers, we obtain the following maps
\begin{align}
   t'&= \frac{1}{2} \left(\frac{\alpha ^2}{x  -t  }+t  +x  \right)\\
   x'&=\frac{1}{2} \left(-\frac{\alpha ^2}{x  -t  }+t  +x  \right)\ ,
\end{align}
 connecting Milne and Rindler observers.

\section{Conformal quantum mechanics}\label{Rstates}
In this section we review the standard construction \cite{deAlfaro:1976vlx,Chamon:2011xk} of conformal quantum mechanics  seen as a one-dimensional conformal field theory.
Let us introduce the generators $H$, $D$ and $K$ as the \emph{quantum mechanical} counterparts of the generators $P_0$, $D_0$ and $K_0$ as $H=i P_0$, $D=i D_0$ and $K=i K_0$ closing the $\mathfrak{sl}(2,\mathbb{R})$ Lie algebra
\begin{equation}
    \comm{H}{D}=i H,\qquad \comm{K}{D}=-i K, \qquad \comm{H}{K}=2 i D\ . 
\end{equation}
The linear combination 
\be\label{quantumham}
G=i \xi = a K + b D + c H\,,
\ee
turns out to be the most general expression for the generator of time evolution in {\it conformal quantum mechanics}, an $\mathrm{SL}(2, \mathbb{R})$-invariant quantum mechanical model introduced and studied extensively in \cite{deAlfaro:1976vlx}. 
The Lagrangian defining the model is given by 
\be
\mathcal{L} = \frac{1}{2}\left(\dot{q}(t)^2 + \frac{g}{q(t)^2} \right)\,,
\ee
where $g$ is a dimensionless coupling constant. The $\mathfrak{sl}(2,\mathbb{R})$ Lie algebra can be realized in terms of the canonically conjugate position and momentum operators $q$ and $p$ as 
\begin{eqnarray}
H = i P_0 &=&  \frac{1}{2}\left(p^2 + \frac{g}{q^2} \right)\\
D = i D_0 &=& t\,H - \frac{1}{4} \left(pq +qp \right)\\
K = i K_0 &=& -t^2\,H +2 t\, D + \frac{1}{2} q^2\ .
\end{eqnarray}
Such model can be understood as a one-dimensional conformal field theory \cite{Chamon:2011xk, Jackiw:2012ur}. Two-point functions are built from the kets $|t\rangle$ satisfying a Schrodinger-like equation
\be\label{tiket}
H |t\rangle = -i \dv{}{t} |t\rangle\,.
\ee
In \cref{rkv} we recalled how the causal character of a radial conformal Killing vector is determined by the sign of the determinant $\Delta=b^2-4ac$. The sign of this quantity also characterizes the conjugacy class to which these generators belong (which determines whether two generators can be mapped into another via an $\mathrm{SL}(2, \mathbb{R})$ transformation) \cite{Camblong:2022oet}: 
\begin{itemize}
     \item $\Delta<0$: \emph{elliptic} generators exhibit a discrete spectrum and they generate a compact subgroup of $\mathrm{SL}(2, \mathbb{R})$. A representative of this class is the generator
\begin{equation}
        R=\frac{1}{2}\left(\alpha H+ \frac{K}{\alpha}\right)
\end{equation}
that, interpreted as a generator of the 3-dimensional Lorentz group of which $\mathrm{SL}(2, \mathbb{R})$ is a double cover, generates rotations in 2+1-dimensional Minkowski space-time.

     \item $\Delta=0$: \emph{parabolic} generators, representatives of this class are $P$ and $K$ and on 2+1-dimensional Minkowski space-time generate {\it null rotations}.
     \item $\Delta>0$: generators of \emph{hyperbolic} transformations; for example  \begin{equation}
         S=\frac{1}{2}\left(\alpha H- \frac{K}{\alpha}\right)\quad\qq{and} \quad D
    \end{equation}
    fall into this class and generate transformations akin to Lorentz boosts seen as generators of the three-dimensional Lorentz group.
\end{itemize}

Let us now go back to the $|t\rangle$ states defined in \eqref{tiket}. The action of the remaining generators of $\mathrm{SL}(2, \mathbb{R})$ on such states is given by \cite{deAlfaro:1976vlx} 
 \begin{align}\label{real1}
     D \ket{t} &= -i \left(t\dv{}{t}+r_0\right)\ket{t}\\
   K \ket{t} &= -i \left(t^2\dv{}{t}+2r_0 t\right)\ket{t}\,.
\end{align}

To study the properties of such states it is convenient to express them in terms of their overlap with states which realize an irreducible representation of the Lie algebra $\mathfrak{sl}(2,\mathbb{R})$ in a discrete basis \cite{deAlfaro:1976vlx,Chamon:2011xk} . In order to describe such representation we define ladder operators
\begin{equation}\label{algebraR}
    L_0= R=\frac{1}{2}\left(\frac{K}{\alpha}+\alpha H\right)\qquad
    L_\pm= S\pm iD = \frac{1}{2}\left(\frac{K}{\alpha}-\alpha H\right)\pm i D\, , 
\end{equation} 
with 
\begin{equation}\label{so21alg}
    \comm{L_0}{L_\pm}=\pm L_\pm, \qquad \comm{L_-}{L_+}=2L_0\,. 
\end{equation}
We define the states $|n\rangle$, labelled by positive integers $n=0,1,...$ via the action of ladder operators

\begin{align}
    L_0\ket{n}&=(n+r_0)\ket{n}\\
    L_\pm\ket{n}&=\sqrt{((n+r_0)(n+r_0\pm 1))-r_0(r_0-1))}\ket{n\pm 1}\, ,
\end{align}
with the normalization
\be\label{innern}
\langle n| n'\rangle = \delta_{nn'}\, .
\ee
The  eigenvalue  $r_0$ of the ground state $|n=0\rangle$ is a positive real number and it is related to the eigenvalue of the Casimir operator
\be
\mathcal{C} = R^2-S^2-D^2 = \frac{1}{2}\left(HK+KH\right)-D^2\,,
\ee
on the $\ket{n}$ states 
\be
\mathcal{C} \ket{n} = r_0(r_0-1) \ket{n}\, .
\ee
For our purposes, we work in the irreducible representation of the Lie algebra $\mathfrak{sl}(2,\mathbb{R})$ belonging to the so-called {\it discrete series} for which $r_0\geq 1$ is integer and half integer. For further details we refer to \cite{Sun:2021thf}. 

From now on we restrict our attention to the case $r_0=1$. In this case, the two-point function of conformal quantum mechanics turns out to have the same functional form of the two-point function of a massless scalar field in Minkowski space-time along the worldline of a static observers sitting at the origin of a spherical coordinate system \cite{Arzano:2021cjm,Arzano:2020thh}. 

We can obtain the overlap between the $|t\rangle$ and $|n\rangle$ states solving the differential equation 
\begin{equation}
    \mel{t}{R}{n}=\frac{i}{2}\left[\left(\alpha+\frac{t^2}{\alpha}\right)\dv{}{t}+2 \frac{t}{\alpha}\right]\ip{t}{n} =(n+1)\ip{t}{n}
\end{equation}
whose solution is given by
\begin{equation}\label{iptn}
  \ip{t}{n}= - \frac{\alpha^2c_n e^{2 i (n+1) \tan ^{-1}\left(\frac{\alpha}{t }\right)}}{\alpha^2+t ^2}\ .
\end{equation}
An analogous differential equation can be written in terms of the ladder operators whose action furnishes the coefficients $c_n$ appearing in the wavefunctions. In fact, the action of $L_+$ leads to
\begin{align}
    \sqrt{(n+1)(n+2)}\ip{t}{n+1}&=\mel{t}{L_+}{n}=\left[\left(i\frac{t}{\alpha}-1\right)+\left(i\frac{t^2}{2\alpha}-i\frac{\alpha}{2}-t\right)\dv{}{t}\right]\ip{t}{n}
\end{align}
giving
\begin{equation}\label{rell+}
    \frac{c_n}{c_{n+1}}=\frac{(n+1) }{\sqrt{(n+1) (n+2)}}
\end{equation}
while the equation for $L_-$ reads
\begin{align}
    \sqrt{n(n+1)}\ip{t}{n-1}&=\mel{t}{L_-}{n}=\left[\left(i\frac{t}{\alpha}+1\right)+\left(i\frac{t^2}{2\alpha}-i\frac{\alpha}{2}+t\right)\dv{}{t}\right]\ip{t}{n}
\end{align}
and one arrives at
\begin{equation}\label{rell-}
    \frac{c_{n-1}}{c_{n}}=\frac{n}{\sqrt{n (n+1)}}\ .
\end{equation}
The general form of the coefficients $c_n$ can thus be obtained from relations \eqref{rell+} and \eqref{rell-} and is
\begin{equation}
    c_n=\sqrt{\frac{\Gamma(2+n)}{\Gamma(n+1)}}=\sqrt{n+1}\ .
\end{equation}
We can now write the $|t\rangle$ states in terms of the  $|n\rangle$ states. Using the relation $2i\tan^{-1}{x}=\log{\frac{1+i x}{1-i x}}$ in \eqref{iptn} and plugging the expression for the coefficients $c_n$ in \eqref{iptn} we obtain
\begin{equation}
\label{tnov}
\ket{t}=\left(\frac{\frac{\alpha+it}{\alpha-it}+1}{2}\right)^{2}\ \sum_{n} (-1)^n \left(\frac{\alpha+it}{\alpha-it}\right)^n\ \sqrt{n+1}\ket{n}\ .
\end{equation}
Such expression can be rewritten in terms of the action of the genertor $L_+$ on the ground state $\ket{n=0}$ 
\begin{equation}
\label{tnzero}
\ket{t}=\left(\frac{\frac{\alpha+it}{\alpha-it}+1}{2}\right)^{2}\ \exp{-\left(\frac{\alpha+it}{\alpha-it}\right)\, L_+} \ket{n=0}\,,
\end{equation}
and in particular, for $t=0$ we obtain the very simple relation
\begin{equation}
\label{tzeron}
\ket{t=0}= \exp\left(- L_+\right) \ket{n=0}\,.
\end{equation}

As discussed in \cite{Chamon:2011xk,Jackiw:2012ur} the inner product between the $|t\rangle$ states can be interpreted as the two-point function of conformal quantum mechanics seen as a one-dimensional CFT. Using the inner product \eqref{innern} from \eqref{tnov} on can obtain
\begin{equation}\label{twpf}
    G(t_1,t_2)\equiv  \braket{t_1}{t_2}=-\frac{ \alpha^{2 }}{4\ (t_1-t_2)^{2 }}\,.
\end{equation}
We notice how this two-point function matches the two-point function of a free massless scalar field in Minkowski space-time along the trajectory of a static inertial observer sitting at $r=0$ apart from a constant factor \cite{Arzano:2020thh}. 

Moreover it can be observed that the $t$-state can be obtained by a complex time translation from the $n=0$ state \cite{Jackiw:2012ur}
\begin{equation}
    \ket{t}= e^{i H t}\ket{t=0}=\frac{1}{4} e^{(\alpha+it) H}\ket{n=0}\ .
\end{equation}
The two-point function \eqref{twpf} can be also written as 
\be\label{2pfvac}
G(t_1,t_2) = \braket{t_1}{t_2} = \bra{t=0} e^{-i H t_1}\, e^{i H t_2}\ket{t=0}\,.
\ee
Let us look at the two-point function \eqref{2pfvac} in terms of eigenstates $\ket{E}$ of the generator $H$. Such $\ket{E}$ states are the conformal quantum mechanical counterpart of the momentum eigenstates $\ket{\mathbf{p}}$ that one introduces in quantum field theory. In fact, one expresses the action of a field operator $\phi(\mathbf{x})$ on the vacuum state in terms of $\ket{\mathbf{p}}$ states
\be\label{phix}
\phi(\mathbf{x})\ket{0}=\int\frac{\dd^3 p}{(2\pi)^3}\frac{1}{2 E_p} e^{-i \mathbf{p} \cdot\mathbf{x} }\ket{\mathbf{p}}\ , 
\ee
where
\be\label{norme2}
\braket{\mathbf{p}}{\mathbf{p}^\prime}=2 E_\mathbf{p} (2\pi)^3\ \delta^{(3)}(\mathbf{p}-\mathbf{p}^\prime)\ 
\ee 
holds. Eigenstates $\ket{E}$ of $H$ are defined by
\be
H\ket{E}=E\ket{E}
\ee
and satisfy the conditions \cite{deAlfaro:1976vlx}
\be\label{norme}
\braket{E}{E^\prime}=\delta (E-E^\prime) \qq{and} \int_0^{+\infty}\ \dd E \ket{E}\bra{E}=\mathbbm{1}\ .  
\ee
The $\ket{t}$ state can be written as the Fourier transform of these states as
\be\label{kettE}
\ket{t}= e^{i H t}\ket{t=0}=\int_{0}^{\infty} \dd E\ \frac{\alpha\sqrt{E}}{2} e^{i E t}\ket{E}\,.
\ee

Comparing \eqref{kettE} and \eqref{phix} we can conclude that the state $\ket{t=0}$ in conformal quantum mechanics plays a role analogous to the inertial vacuum for quantum fields in Minkowski space-time as the averaging state on which one builds the two-point function.\\ 

We now get to a crucial observation. Let us recall that the $\mathfrak{sl}(2,\mathbb{R})$ Lie algebra  \eqref{so21alg} can be realized in terms of two sets of creation and annihilation operators $a^\dagger_L, a^\dagger_R, a_L, a_R$
\begin{equation}\label{algdoppia}
    L_0=\frac{1}{2}\left(a^\dagger_L a_L+a^\dagger_R a_R+1\right)\ , \quad L_+=a^\dagger_L a^\dagger_R \qq{and} L_-=a_L a_R\,.
\end{equation}
This shows that the ground state of the $L_0$-operator has a {\it bipartite structure}
\begin{equation}
    \ket{n=0}=\ket{0}_L \otimes \ket{0}_R\,,
\end{equation}
and that the $\ket{t=0}$ state in \eqref{tzeron} can be written as 
\begin{equation}
\label{tvac}
    \ket{t=0}=e^{-a^\dagger_L a^\dagger_R} \ket{0}_L \ket{0}_R=\sum_n\ (-1)^n \ket{n}_L \ket{n}_R = -i 
    \sum_n\ e^{i \pi L_0} \ket{n}_L \ket{n}_R \,.
\end{equation}
From the last equality it is clear that the $\ket{t=0}$ state exhibits a structure similar to that of a {\it thermofield double state}\footnote{Such state can be built by ``doubling" the oscillator's degrees of freedom (see \cite{Lykken:2020xtx} for a review) and is defined by the superposition
\be\label{tfd1}
|TFD\rangle = \frac{1}{Z(\beta)} \sum^{\infty}_{n=0} e^{-\beta E_n/2} |n\rangle_L \otimes |n\rangle_R\,,
\ee
where $Z(\beta) = \sum^\infty_{n=0} e^{-\beta E_n}$ is the partition function at inverse temperature $\beta$. The state \eqref{tfd1} is highly entangled and, tracing over the degrees of freedom of one copy of the system, we obtain a thermal density matrix
\be
Tr_L\{|TFD\rangle \langle TFD|\} =\frac{e^{-\beta \mathcal{H}}}{Z(\beta)} 
\ee
at a temperature $T=1/\beta$. The Hamiltonian $\mathcal{H}$ is known as {\it modular Hamiltonian}.} for the Hamiltonian $-iL_0$. As we will discuss in detail in the coming sections, there are alternative realizations of the algebra of ladder operators \eqref{algdoppia} for which $-iL_0 = D$ or $-iL_0 = S$. We thus have, in analogy with quantum field theory in Minowski space-time that, as outlined above, the state $\ket{t=0}$ is the analogue of the inertial or {\it Hartle-Hawking} state while the state $\ket{n=0}$ plays a role similar to the {\it Boulware vacuum} exhibiting a bi-partite structure. As for their higher-dimensional analogues, the Hartle-Hawking state $\ket{t=0}$ exhibits the structure of a thermofield double state built on the bi-partite Boulware vacuum $\ket{n=0}$ populated by excitations of the modular Hamiltonians $-iL_0 = D$ or $-iL_0 = S$ which generate time evolution only on a portion of the geometric domains of the theory. In the following sections we will develop the details of the identification of $iL_0$ with the Hamiltonians $D$ and $S$.

\section{Hyperbolic generators as modular Hamiltonians}
Following the observations above, in this section we discuss the realization of the $\mathfrak{so}(2,1)$ Lie algebra with $-iL_0$ identified with the hyperbolic generators $S$ and $D$ and study their eigenfunctions. 

\subsection{Eigenstates of the $iS$-generator}
We start by focusing the attention on the hyperbolic generator $S$.  We note that from the definition of the generator of rotations $R=\frac{1}{2}\left(\frac{K}{\alpha}+\alpha H\right)$ we obtain $i S$ by mapping $\alpha\mapsto i \alpha$. Applying the map $\alpha\mapsto i\alpha$ to \eqref{algebraR} we obtain another $\mathfrak{so}(2,1)$ algebra with the following set of generators
\begin{equation}
\label{algebraS}
    L_0=iS\ ,\qquad L_+=i(D-R)\ ,\qquad L_-=-i(D+R)\ .
\end{equation}
Seemingly, we expect that the states constructed starting from the algebra of \eqref{algebraS} are mapped to the $t$-states constructed from the algebra in which $L_0=R$. 
We can verify this by repeating the same analysis carried in \cref{Rstates}. In this case, the differential equation to write becomes \cite{Arzano:2023pnf}
\begin{equation}
    (n+1)\, {}_S\ip{t}{n}= {}_S\mel{t}{i S}{n}=-\frac{1}{2}\left[\left(\alpha-\frac{t^2}{\alpha}\right)\dv{}{t}-2 \frac{t}{\alpha}\right] {}_S\ip{t}{n}
\end{equation}
which is solved for
\begin{equation}
 {}_S \ip{t}{n}= c_n\  \frac{(t-\alpha)^n a^2}{(t+\alpha)^{n+2}}\ .
\end{equation}
Again, we act with the ladder operator $L_+$ 
\begin{align}
    \sqrt{(n+1)(n+2)}\, {}_S\ip{t}{n+1}&=\, {}_S\mel{t}{L_+}{n}=\left[\left(-t+\frac{1}{2}\left(\frac{t^2}{\alpha}+\alpha\right)\right)\dv{}{t}+\left(-1+2\frac{t}{\alpha}\right)\right] {}_S \ip{t}{n}
\end{align}
and obtain
\begin{equation}
  c_n=c_{n+1} \sqrt{\frac{n+1}{n+2}}\ ,
\end{equation}
 while the action of $L_-$ gives
\begin{align}
    \sqrt{n(n+1)}\, {}_S\ip{t}{n-1}&= {}_S\mel{t}{L_-}{n}=\left[\left(t-\frac{1}{2}\left(\frac{t^2}{\alpha}+\alpha\right)\right)\dv{}{t}+\left(1+2\frac{t}{\alpha}\right)\right]  {}_S\ip{t}{n}
\end{align}
whose solution reads
\begin{equation}
    c_n=  c_{n-1} \sqrt{\frac{(n+1)}{n}}.
\end{equation}
By combining together the results of the action of the ladder operators we conclude that
\begin{equation}
    c_n= \sqrt{\frac{\Gamma(2+n)}{n!}}\ ,
\end{equation}
in terms of which the $t$- state is
\begin{equation}\label{tauis}
\begin{split}
    \ket{t}_S&=\frac{\alpha^2}{(\alpha+t)^2}\sum_n  \sqrt{\frac{\Gamma(2+n)}{n!}} \left(\frac{t-\alpha}{\tau+\alpha}\right)^n\ket{n}= \frac{\alpha^2}{(\alpha+t)^2}\sum_n \left(\frac{t-\alpha}{t+\alpha}\right)^n \frac{L_+^n}{n!}\ket{n=0}\\&=\frac{\alpha^2}{(\alpha+t)^2} e^{\frac{t-\alpha}{t+\alpha} L_+}\ket{n=0}\ .  
\end{split}
\end{equation}

As expected, the result \eqref{tauis} coincides with the $t$-state \eqref{tnov} by mapping $\alpha\mapsto i \alpha$. Additionally, with simple manipulations, we observe that the inertial vacuum
\begin{equation}
    \ket{t=0}_S = -i 
    \sum_n\ e^{- \pi S} \ket{n}_L \ket{n}_R \,
\end{equation}
has the structure of a thermofield double state with temperature
\begin{equation}\label{Ts}
    T_S=\frac{1}{2\pi\alpha}
\end{equation}
for the modular Hamiltonian $S/\alpha$ generating diamond time evolution. Notice that \eqref{Ts} is just the diamond temperature for diamond observers sitting at the origin \cite{Martinetti:2002sz, Alonso-Serrano:2020pcz, Alonso-Serrano:2021eju, Chakraborty:2022qdr,Chakraborty:2023fsz}.

\subsection{Eigenstates of the $iD$-generator}
\label{l0id}
As already discussed in \cref{diam}, we can map the $iS$ generator into the generator $iD$ via the coordinate transformation 
 \begin{equation}\label{mappasd}
    t^\prime=\frac{\alpha(t-\alpha)}{\alpha+t}\,.
 \end{equation}
Applying this map to the algebra \eqref{algebraS} we obtain another set of operators closing the $\mathfrak{so}(2,1)$ 
\begin{equation}
\label{algebraD}
    L_0=iD\ ,\qquad L_+=-i \alpha H\ ,\qquad L_-=-i \frac{K}{\alpha}\ .
\end{equation}
For the eigenfunctions of the operator $i D$ we look for solutions of the differential equation \cite{Arzano:2023pnf}
\begin{equation}
    (n+1)\,{}_D\ip{t}{n}={}_D\mel{t}{i D}{n}=-\left[t\dv{}{t}+1\right]{}_D\ip{t}{n}
\end{equation}
which are given by
\begin{equation}
  {}_D\ip{t}{n}= c_n\  t^{-n-2}\ .
\end{equation}
The coefficients $c_n$ are again obtained by recursion via the action of the ladder operators. Starting with $L_+$ 
\begin{align}
    \sqrt{(n+1)(n+2)}\,{}_D\ip{t}{n+1}&={}_D\mel{t}{L_+}{n}= \alpha \dv{}{t}{}_D\ip{t}{n}
\end{align}
we obtain
\begin{equation}
    \sqrt{(n+1)(n+2)}\ c_{n+1}=- \alpha\ c_n (2+n)\ ,
\end{equation}
while the action with $L_-$ gives
\begin{align}
    \sqrt{n(n+1)}\,{}_D\ip{t}{n-1}&={}_D\mel{\tau}{L_-}{n}=\frac{1}{\alpha}\left[t^2 \dv{}{t}+2t\right]{}_D\ip{t}{n}
\end{align}
whose solution is
\begin{equation}
    - \alpha \sqrt{n(n+1)} c_{n-1}=n\  c_n\ .
\end{equation}
By combining the two results we arrive at the following expression for the $c_n$
\begin{equation}
    c_n= \frac{(-1)^n}{2}\ \alpha^{n+2} \sqrt{\frac{\Gamma (2+n)}{n!}}
\end{equation}
in terms of which the $t$-state becomes
\begin{equation}\label{tauid}
\begin{split}
    \ket{t}_D&=\frac{1}{2}\left(\frac{\alpha}{t}\right)^2 \sum_{n} (-1)^n \left(\frac{\alpha}{t}\right)^n \sqrt{\frac{\Gamma (2+n)}{n!}} \ket{n}=\frac{1}{2}\left(\frac{\alpha}{t}\right)^2 \sum_{n} (-1)^n \left(\frac{\alpha}{t}\right)^n \frac{L_+^n}{n!} \ket{n=0}\\&=\frac{1}{2}\left(\frac{\alpha}{t}\right)^2 e^{- \frac{\alpha}{t}L_+}\ket{n=0}\ .    
\end{split}
\end{equation}
According to our expectation, this state can be obtained from \eqref{tauis} using the map \eqref{mappasd}. Notice that this state is ill defined at $t=0$ coherently to the fact that the $D$ generator is ill defined at that point. We see that now the state $t=\alpha$ exhibits the structure of a thermofield double state for the modular Hamiltonian $D/\alpha$ at the temperature $T_D=1/(2\pi \alpha)$. The point $t=\alpha$ is the image of the origin under the conformal mapping \eqref{mappasd} and it corresponds to the origin of the conformal time $\tau$ variable defined by $t = \alpha\ e^{ \frac{\tau}{\alpha}}$. 
\\

In the cases explored throughout this section we have noticed that the identification \eqref{algebraS} can be obtained by ``Wick rotating" the length parameter $\alpha \rightarrow i \alpha$. Under this map the generator $R$ turns into $iS$ and the wavefunctions \eqref{iptn} into eigenfunctions of the operator $L_0=iS$. Following steps analogous to the ones leading to \eqref{tnzero} we found the state $\ket{t}$ as a function of a tower of eigenstates of the operator $i S$. Additionally, one can further obtain the wavefunctions of the operator $iD$ applying to the latter states the map \eqref{mappasd} which indeed allows to write $i S$ as $i D$. Our arguments are supported by the observation that $L_0$ can be identified with the elliptic generator $R$ of the $SO(2)$ compact subgroup of $SL(2,\mathbb{R})$ \cite{deAlfaro:1976vlx,Chamon:2011xk,Jackiw:2012ur} while its ``Wick-rotated" counterpart $i L_0$ will generate non-compact transformations thus we are led to identify it with an hyperbolic generator. In fact, one finds \cite{Arzano:2021cjm} that the generators of the Lie algebra \eqref{so21alg}, besides the identification \eqref{algebraR}, have two alternative realizations in terms of the generators $H,D$ and $K$ given in \eqref{algebraS} and \eqref{algebraD}.\\

As illustrated by eq. \eqref{tvac}, the state $\ket{t=0}$ is an entangled state in relation to the bi-partition of the Hilbert space based on L and R degrees of freedom \cite{Arzano:2023pnf}. Drawing a parallel with the inertial vacuum depicted as a thermofeld double state over the left and right Rindler modes excitations (the two corresponding domains of the boost modular Hamiltonian's evolution) \cite{Valdivia-Mera:2020nko}, we can interpret the two sets of degrees of freedom in \eqref{tvac} as part of the diamond and Milne time evolution domains and their complements (our result corresponds to the restriction to $r = 0$ worldlines of the entanglement examined in \cite{Higuchi:2017gcd} and \cite{Olson:2010jy}). We can measure this entanglement by computing the Von Neumann entropy of the reduced density matrix obtained by tracing over one set of degrees of freedom in the density matrix representing the inertial vacuum state $\ket{t=0}$. This entanglement entropy can be viewed as the $0 + 1$-dimensional equivalent of the entanglement entropy of a quantum field across space-time regions. Contrary to its higher dimensional equivalents, the simplicity of the model renders the computation of the entanglement entropy rather straightforward as we discuss in detail in the next section.

\section{Entanglement entropy}
To obtain the entanglement entropy associated to the partition of the $\ket{t=0}$ state, we should first observe that this state cannot be normalized. This is expected in light of the identification of the inner product $\braket{t_1}{t_2}$ with  the restriction of the two-point function of a massless field to the $r=0$ worldline and its divergence at coincident points.

We regularize such UV divergence via an infinitesimal translation in imaginary time. We thus consider the state 
\begin{equation}
\label{eq2}
\ket{t=i\epsilon}=\left(\frac{\frac{\alpha-\epsilon}{\alpha+\epsilon}+1}{2}\right)^{2}\ e^{-\frac{\alpha-\epsilon}{\alpha+\epsilon}\ L_+}\ket{n=0}\ .
\end{equation}
This regularization consists in introducing a short-distance cut-off scale, $\epsilon$. We can rewrite \eqref{eq2} as
\begin{equation}
    \ket{t=i\epsilon}=\left(\frac{\alpha}{\alpha+\epsilon}\right)^{2} \sum_{n=0}^{\infty}(-1)^n\left(\frac{\alpha-\epsilon}{\alpha+\epsilon}\right)^n \ket{n}_L\ket{n}_R\ .
\end{equation}
From
\begin{equation}
    \bra{t=i\epsilon}\ket{t=i\epsilon}=\frac{\alpha }{4\epsilon}\left(\frac{\alpha}{\alpha+\epsilon}\right)^{2}\equiv \frac{1}{\mathcal{N}^2}\ 
\end{equation}
we normalize the $\ket{t=i\epsilon}$ state and introduce a state $\ket{\delta}$ with unitary norm
\begin{equation}
    \ket{\delta} \equiv \mathcal{N} \ket{i\epsilon}\ .
\end{equation}
We now consider the density matrix 
\begin{equation}
\label{rhoLR}
    \rho_{RL}=\ket{\delta}\bra{\delta}
\end{equation}
and obtain the reduced density matrix $\rho_L$ by tracing over the R degrees of freedom
\begin{equation}
\label{rhoL}
    \rho_L=\Tr_R{\rho_{LR}}=\mathcal{N}^2 \left(\frac{\alpha}{\alpha+\epsilon}\right)^4 \sum_{n=0}^\infty \left(\frac{\alpha-\epsilon}{\alpha+\epsilon}\right)^{2n} \ket{n}_L\bra{n}_L \ .
\end{equation}
The Von Neumann entropy of such reduced density matrix is then given by
\begin{equation}
    S=-\Tr{\rho_L\log\rho_L}=-\frac{(\alpha-\epsilon )^2 \log \left(\frac{(\alpha-\epsilon )^2}{(\alpha+\epsilon )^2}\right)}{4 \alpha \epsilon }-\log \left(\frac{4 \alpha \epsilon }{(\alpha+\epsilon )^2}\right)\,,
\end{equation}
from which, performing the limit $\epsilon \rightarrow 0$, we arrive at
\begin{equation}
\label{nostro}
    S=\log \left(\frac{\alpha}{\epsilon}\right)+\text{const}+\order{\epsilon^2}\, .
\end{equation}
We observe that the result obtained is logarithmic divergent when the UV cut-off scale $\epsilon$ is sent to zero. This is a known feature in quantum field theory where in $d$-dimensional space-time the entanglement entropy associated to a spatial region $\mathcal{A}$ is proportional to
\be
\frac{\text{Area}(\partial \mathcal{A})}{\epsilon^{d-2}}\ ,
\ee
with $\text{Area}(\partial \mathcal{A})$ being the area of the boundary of the region (entangling surface) while $\epsilon$ is a UV cut-off. Precisely, $S_{\mathcal{A}}$ has the general expressions \cite{Rangamani:2016dms}
\begin{align}
   S_\mathcal{A}&=g_{d-2} \left(\frac{\alpha}{\epsilon}\right)^{d-2}+g_{d-4} \left(\frac{\alpha}{\epsilon}\right)^{d-4}+\dots +g_1 \left(\frac{\alpha}{\epsilon}\right)+ (-1)^{\frac{d-1}{2}} g_0+ \mathcal{O}(\epsilon)\qquad  &\qq{d odd}\\\label{entropycft}
   S_\mathcal{A}&= g_{d-2} \left(\frac{\alpha}{\epsilon}\right)^{d-2}+g_{d-4} \left(\frac{\alpha}{\epsilon}\right)^{d-4}+\dots+ (-1)^{\frac{d-2}{2}} g_0\log\left(\frac{\alpha}{\epsilon}\right)+ \mathcal{O}(\epsilon^0)  \qquad &\qq{d even}
\end{align}
where $g_i$ with $i>0$ depend on the regularization scheme, $\alpha$ is linked to the size of the region $\mathcal{A}$ and $\epsilon$ is again a UV regulator. The coefficient $g_0$ is particularly important since it carries non-trivial and universal information. 
Two-dimensional conformal field theories exhibit a logarithmic divergence in the entanglement entropy. For example if we consider the entanglement between a shorter line-segment with length $\alpha$ and a longer one with length $L$ containing it in the limit  $\frac{\alpha}{L}\ll 1$, we have \cite{Calabrese:2004eu,Holzhey:1994we,Saravani:2013nwa,Rangamani:2016dms}
\begin{equation}
\label{rid}
    S=\frac{c}{3}\log{\frac{\alpha}{\epsilon}}+\text{const}\ ,
\end{equation}
where $c$ is the central charge of the CFT  and it is equal to $1$ in a quantum field theory of a massless bosonic field. Heuristically, such logarithmic divergence can be seen as a limiting case of the power law divergence and it is in agreement with the scenario where the entangling surface is built up by a set of disconnected points. It is noteworthy that the behaviour of our result \eqref{nostro} for the entanglement entropy in conformal quantum mechanics coincides with the one in found in two-dimensional conformal field theory. In particular, it exhibits the same logarithmic divergence, in agreement with the point-like nature of the entangling surface.

\section{Conclusions}
In our contribution, we evidenced the capacity of conformal quantum mechanics, a simple one-dimensional quantum mechanical model, to replicate thermal phenomena associated to the freedom in the choice of time evolution which one encounters in quantum field theory in the presence of causal horizons. The existence of a correspondence between radial conformal Killing vectors in 3+1-dimensional Minkowski space-time and the generators of time evolution in conformal quantum mechanics, which we reviewed in \cref{rkv}, is suggestive of this fact. Conformal Killing horizons in Minkowski space-time correspond to boundaries in the domain of time evolution in conformal quantum mechanics. As in quantum field theory in Minkowski space-time, we were able to identify states labelled by a globally defined time-variable in conformal quantum mechanics which exhibit a natural bipartite structure in terms of excitations of a hyperbolic Hamiltonian. When restricted to one set of such excitations, the state we identified as {\it inertial vacuum} is described by a thermal density matrix with temperatures analogous to the diamond and Milne ones in Minkowski space-time. These findings offer group-theoretic support for the existence of Milne and diamond temperatures. Furthermore, our model allows a relatively straightforward computation of the entanglement entropy for such an inertial vacuum, a notoriously challenging task in higher-dimensional quantum field theories.
Since the boundaries of the time domains encompassed by the hyperbolic Hamiltonians are evidently point-like, the entanglement entropy we derive exhibits a logarithmic divergence akin to that observed in two-dimensional conformal field theories, where the boundaries of the spatial region under consideration are also point-like.

Our findings also suggest a profound connection between the thermodynamic properties of causal diamonds and of the Milne patch of Minkowski space-time. The correspondence with states in conformal quantum mechanics may furnish novel, robust group-theoretic tools for exploring the properties of entanglement across the boundaries of domains in Minkowski space-time. We defer the exploration of this intriguing avenue to future investigations.

\providecommand{\href}[2]{#2}\begingroup\raggedright
\end{document}